\documentclass[12pt]{article}

\usepackage{graphics}
\usepackage{amsmath}
\def\ni{\noindent}
\def\ph{{\phantom{...}}}
\def\={\phantom{..} = \phantom{..}}

\def\+{\phantom{..} + \phantom{..}}
\def\>{\phantom{..} > \phantom{..}}
\def\<{\phantom{..} < \phantom{..}}
\def\-{\phantom{..} - \phantom{..}}

\def\bq{\begin{quote}}
\def\eq{\end{quote}}
\def\be{\begin{equation}}
\def\ee{\end{equation}}
\def\bar{\begin{eqnarray}}
\def\ear{\end{eqnarray}}
\def\no{\nonumber}

\def\half{{\frac{1}{2}}}

\def\Sch{Schr{\"o}dinger}
\def\Schism{Schr{\"o}dingerism}
\def\Schist{Schr{\"o}dingerist}
\def\Schists{Schr{\"o}dingerists}

\def\Schseqn{Schr{\"o}dinger's equation}

\def\Copism{Copenhagenism}
\def\Copist{Copenhagenist}
\def\Copists{Copenhagenists}

\def\wf{wavefunction}
\def\wfs{wavefunctions}
\def\Re{\hbox{Re}}

\def\MP{Measurement Problem}

\def\BM{Brownian Motion}
\def\OU{Ornstein-Uhlenbeck process}
\def\qp{quasi-periodic}

\def\cH{{\cal H}}
\def\Emax{E_{\hbox{max.}}}

\def\sumn{{\sum_{n=1}^N}}
\def\sumk{{\sum_{k=1}^{\infty}}}
\def\sumninf{{\sum_{n=1}^{\infty}}}

\def\Re{\hbox{Re}}

\def\cE{{\cal{E}}}
\def\BML{Brownian-Motion-Like}

\title{\bf Can \Schist\ Wavefunction Physics Explain \BM?   \\[2in]
       }

\author{W. David Wick\footnote{email: wdavid.wick@gmail.com}}

\begin{document}
\maketitle
\pagebreak

\section*{Abstract}

Einstein's 1905 analysis of the Brownian Motion of a pollen grain in a water droplet
as due to statistical variations
in the collisions of water molecules with the grain, 
followed up by Perrin's experiments, 
provided one of the most convincing demonstrations of the reality of atoms. 
But in 1926 \Sch\ replaced classical particles 
by wavefunctions, 
which cannot undergo collisions. 
Can a \Schist\ wavefunction physics account for Perrin's observations? 
As systems confined to a finite box
can only generate quasiperiodic signals, this seems impossible, 
but I argue here that the issue is more subtle.
I then introduce several models of the droplet-plus-grain; unfortunately,
no explicit solutions are available (related is the
remarkable fact that the harmonics of a general right trangle are still unknown).
But from generic features of the models I conclude that: (a) wavefunction models may generate 
trajectories resembling those of a stochastic process; (b) diffusive behavior
may appear for a restricted time interval; and (c) additional ``Wave Function Energy", by
restricting ``cat" formation, can render the observations more ``classical".  
But completing the Einstein program of linking diffusion to viscosity and temperature 
in wavefunction models is still challenging.

\pagebreak

\section{Introduction}

In 1905, Einstein 
proposed,\footnote{But Henri Poincar{\'e}
got there first, lecturing on \BM\ in 1900 at the Paris jubilee. He gave the equations
to his student Bachilier, but the latter was not interested 
in pollen grains and water molecules but in stock markets, and subsequently 
published the first serious work in mathematical finance.}\cite{einstein}, 
that a pollen grain in a drop of water, under bombardment by
water molecules, would undergo a kind of continuous version of probability
 theory's ``drunkard's walk". 
(Einstein remarks, in passing, that he does not know whether his theory 
related to the observations of Robert Brown, published in 1828.)
To justify the
random part, he imagined that a group of water molecules striking the grain in some time
period would be replaced, after a lapse of time, by another group. If the first group
transmitted a net impulse to the grain in some direction (due to statistical variation 
in their velocities), the second group would also,
but the impulses would be independent. Hence the random walk. (Later, mathematicians
including Wiener, Ornstein, and Uhlenbeck refined this proposal into a stochastic process
with decay of correlations over time.) Einstein also linked the temperature 
and the viscosity of the water to the `diffusion constant', 
which permitted a calculation of Avogadro's Number
using only a thermometer, a stopwatch, a ruler, and a microscope (plus some pollen grains),
from which Perrin, \cite{perrin}, later derived a value compatible with other determinations
from chemistry and electrical theory. All together, these developments represented
a refutation of the positivist doubts about the existence of atoms (and won Perrin a Nobel Prize;
Einstein got it for his `light quantum' but probably deserved one also for his \BM\ paper.)

Our problem post-1920s is that we no longer believe in classical particles but in 
``quantum states", which for we \Schists\ are wavefunctions. Wavefunctions cannot
undergo collisions or exhibit collisions. So can the \Schist\ program hope to explain 
\BM?

To have a concrete model for discussion, we might imagine a single heavy particle
surrounded by light particles, all confined to a finite box, and 
the whole system described by a wavefunction. 
Of course, we will be interested, perversely, in the motion of the 
heavy particle rather than in the light ones, the more 
usual preoccupation in scattering theory. 
The conceptual difficulty now becomes apparent: 
for any observable of the system, say $<\psi(t)|X|\psi(t)>$, where $\psi(t)$
denotes the \wf\ at time `$t$' and $X$ denotes the heavy-particle position,
we will find an expression of form:

\be
f(t) \= \Re\,\sumk\,a_k\,\exp\{\,i\nu_k\,t\,\},\label{qpdef}
\ee

\ni where `$i$' denotes $\sqrt{-1}$, the $a_n$ are (possibly complex) constants,
 and the $\nu_k$
are real frequencies (which of course are energies, divided by $\hbar$),
and I have assumed the usual, linear quantum evolution law. (Elsewhere, I proposed
a nonlinear modification of \Sch's equation as a solution to the \MP, \cite{WickI}, but I delay
raising that issue until we decide what conventional theory provides.)
Thus, at least for a system in a finite box, all that conventional theory can
produce for observables is a ``\qp" signal.

So the issue becomes: can a \qp\ signal resemble, 
in some sense, a trajectory of a stochastic process? Can it ever be said to
be ``\BML" (BML)?

Let me clarify the meaning of that last phrase, 
interpreted scientifically rather than mathematically.
\BM\ is the name we give to the phenomenon observed by Brown, Perrin, and others. 
The Wiener and OU processes are models. 
Mathematicians who wish to contribute to a scientific field sometimes 
make a mistake, usually after reading some exposition by a fellow mathematician and imagining that
the model presented there defines the subject of study.
Thus a mathematician asked ``what is \BM?" is quite likely to emphasize the
independence  of increments in different time intervals
assumed by Wiener. But I doubt that any experimentalist ever acquired 
enough measurments on a single pollen grain (or whatever was observed) 
to apply any statistical tests for independence.  

We must not confuse phenomena with models.
So what does real \BM\ look like,
according to Perrin, \cite{perrin}? 
First, the motion of the pollen grain appears irregular, like a random walk, with many
apparent reversals of direction. Let's call such trajectories `\BML'. 
Second, the motion of many such grains does not appear 
ballistic, traversing equal distances with equal times 
(as they would if swept along by eddy currents in the droplet), but rather diffusive, traversing
distances that grow only with the square-root of time. Can our formula exhibit these 
behaviors---'\BML' and/or Diffusive? 
Before introducing a class of \wf\ models, I address the first question in the next section. 

I note that there is a large literature accesible under the search phrase:
``quantum Brownian motion".\footnote{On 4/26/2023, at 12:06, PST, 
a Google search on the phrase yielded
1.67 million hits.} 
But I will not be quoting results or deriving techniques from it.
The dominant philosophy guiding these papers is that of Born, Heisenberg, Bohr,
von Neumann and others of the \Copist\ camp. E.g., quantities like 
$<\psi(t)|A|\psi(t)>$ are interpreted as ``the expected value of (operator) $A$ in state $\psi(t)$",
which \Schists\ reject, and projections (``collapses") are added by theorists' {\em fiat}.
Connecting derived formulas to, e.g., Perrin's observations, are lacking, usually because the
systems considered are very unlike a nearly-macroscopic pollen grain in a drop of water.
 Moreover, the methodology typically involves many substitutions or articulations,
e.g., replacing \Schseqn\ by a functional integral, or a master equation, 
or adding stochastic terms. Questionable approximations or truncations abound. 
For a review, see 
H{\"a}nggi and Ingold
2004,
\cite{HandI}, who concluded: 
``... the topic of a quantum Brownian motion ... cannot be considered as `solved'."

\section{Part One: Can a Quasiperiodic Signal Appear `\BML'?}

\subsection{Wiener's formula and the \OU\label{formula_sect}}

At first sight, the \qp\ function defined in (\ref{qpdef}) seems almost the antithesis
of a random trajectory sampled from a genuine stochastic process, 
such as Wiener's or the OU process. But appearances can be deceiving.

In the 1920s, Wiener formulated a continuous-time version of probability theory's drunkard's walk,
usually written $w(t)$, having the properties: for any collection of times $\{t_j\}$ the
$\{w(t_j)\}$ are (jointly) Gaussian, mean-zero random variables; disjoint intervals yield 
independent increments, e.g., $w(u) - w(t)$ and $w(s) - w(r)$ are independent if
the intervals $[t,u]$ and $[r,s]$ don't overlap; and 

\be
\cE\,w(t)^2 \= t.
\ee

\ni where $\cE$ denotes expectation.
 
Wiener's constructed his process {\em via} random Fourier series, as follows.
Consider the complex process:

\bar
\no w_c(t) &\=&  \int_0^t\,ds\,\left\{\,a_o + \sumninf\,a_n\,\exp(i\,\pi n\,s)\,\right\} \\
\no &\=& a_0\,t + \sumninf\,a_n\,\frac{1}{i\pi n}\,\left\{\,\exp(i\pi n\,t) - 1\,\right\}.\\
&&\label{weinerform}
\ear

Let the $\{a_n\}_0^{\infty}$ be selected as i.i.d. N(0,1) (real) random variables.
I claim that:

\be
\cE\,|w_c(t)|^2 
\= t.
\ee

The proof relies on a well-known fact about Fourier series.
Consider the Fourier coefficients of the indicator function of the interval [0,t]:

\be
\int_0^1\,ds\,1_{[0,t]}(s)\,\exp(i \pi n s) \= \frac{1}{i \pi n}\left\{\,
\exp(i \pi n t) - 1\,\right\},
\ee

\ni for $n \geq 1$ and $t$ for the case $n=0$.
Then since forming Fourier coefficients yields an isometry from
L$^2$[0,1] into l$^2$ we see that 

\bar
\no && \cE\,|w_c(t)|^2 
\= t^2 + \sumninf\,\frac{1}{(\pi n)^2}\,|\,\exp(i\pi n\,t) - 1\,|^2 \=\\
\no && \int_0^1\,ds 1_{[0,t]}^2\, \= t,\\
&&
\ear

\ni proving the claim for $t \leq 1$ (longer intervals can be accomodated by scaling). 

(I note for later reference that the only facts about the $\{a_n\}$
used in the above calculation are that they have a second moment,
the marginal distributions are identical,
and $\cE\,a_n\,a_m = 0$ for $n \neq m$. For example, this will be true if the sequence is chosen
from the uniform distribution on the infinite-dimensional sphere defined by $\sum a_n^2 = 1$,
if it can be properly defined.)

The real part of (\ref{weinerform}) is:

\be
\hbox{Re}\,w_c(t) \= a_o\,t \+ \sumninf\,a_n\,\frac{\sin(\pi n t)}{\pi n},\label{weiner}
\ee

\ni a formula that is found in various sources.\footnote{The Wikipedia page on Wiener's process
gives it with a factor of $\sqrt{2}$ before the second term. Maybe that comes from
normalization in the isometry theorem? Or their definition of $w(t)$?}

Many physicists do not accept the Wiener process as a model of \BM, because
(as is well-known) 
the Wiener trajectories are nowhere differentiable (although continuous) in time.
Hence our pollen grain would not be allowed a velocity. So they prefer the \OU\
as the more-plausible model of the phenomenon.

The \OU\ is 
defined\footnote{In many places the ``\OU" is taken to be just
the velocity part defined here; but we are interested in positions.} 
(informally) by:

\be
\frac{dv}{dt} \= - \gamma\,v \+ \frac{dw}{dt}.
\ee

\ni where $v$ is interpreted as velocity, $\gamma$ is a positive constant, and 
for the position $x$:

\be
x(t) = x(0) + \int_0^t\,v(s)\,ds.
\ee 

By writing where it appears:

\be
\frac{dw}{dt}\,dt = dw(t),
\ee

and exploiting the assumed properties of Weiner's differential:

\be
\cE\,dw(s)\,dw(t) = \delta(t-s),
\ee

\ni we can derive the formula with a ``Weiner integral":

\be
v(t) = \exp(-\gamma\,t)\,v(0) + \int_0^t\,\exp(-\gamma [t - s]\,)\,dw(s).
\ee

A lengthy calculation (left to the reader) assuming $v(0) = 0$ 
then gives for the mean-squared displacement:

\be
\cE\,\left[\,x(t) - x(0)\,\right]^2 \= \frac{1}{\gamma^2}\,\left\{\,t + \frac{1}{2\gamma}\,
\left[\,1 - \exp(-2\gamma t)\,\right] - \frac{2}{\gamma}\,\left[\,1 - \exp(-\gamma t)\,\right]
\,\right\}.\label{OUMSdisplacement}
\ee

From this we see that, at long times or if $\gamma$ is large (with the interpretation
that ``friction" is large and the velocity essentially follows the Weiner process) 
there is left a term proportional to `$t$'.

Can the \OU\ be obtained from a formula, as (\ref{weiner}) for Weiner's process?
Let's write, formally,

\be
\frac{dw}{dt} \= \sumninf\,a_n\,\exp(\,\pi i n t\,) \+ a_0,
\ee

\ni from which, assuming $v(0) = 0$, we obtain:

\bar
\no v(t) &\=& \int_0^t\,ds\,\exp(-\gamma [t - s]\,)\,
\left\{\,\sumninf\,a_n\,\exp(\pi i n s) \+ a_0\,\right\} \\
\no &\=& \sumninf\,a_n\,\left\{\frac{1}{\gamma + \pi i n}\,\right\}\,\exp(\pi i n t) \+\\
\no && \frac{a_0}{\gamma} - \exp(-\gamma t)\,\left\{\, \frac{a_0}{\gamma}
 +  \sumninf\,a_n\,\left\{\frac{1}{\gamma + \pi i n}\,\right\}\,\right\}.\\
&&
\ear

\ni Hence, assuming also $x_c(0)=0$:

\bar
\no x_c(t) &\=& \int_0^t\,v(s)\,ds\\
\no &\=& \sumninf\,a_n\,\left\{\frac{1}{\gamma + \pi i n}\,\right\}\,
\left\{\frac{1}{\pi i n}\,\right\}\,
\left\{\,\exp(\pi i n t) - 1\,\right\} \+\\
\no && \frac{a_0 t}{\gamma} - \frac{1}{\gamma}\,[\,1 - \exp(-\gamma t)\,]
\left\{\,\sumninf\,a_n\,\left\{\frac{1}{\gamma + \pi i n}\,\right\} 
+ \frac{a_0 }{\gamma}\,\right\}.\\
&&
\ear

\ni Taking the real part (assuming real $a_n$):

\bar
\no \hbox{Re}\,x_c(t) &\=& \\
\no && \sumninf\,a_n\,\left\{\frac{\gamma}{\gamma^2 + \pi^2 n^2}\,\right\} 
\,\left\{\frac{1}{\pi n}\,\right\}\,\sin(\pi n t) \+\\
\no && \sumninf\,a_n\,\left\{\frac{1}{\gamma^2 + \pi^2 n^2}\,\right\} 
\,\left\{\, 1 - \cos(\pi n t)\,\right\} \+\\
\no && \frac{a_0 t}{\gamma} - \frac{1}{\gamma}\,[\,1 - \exp(-\gamma t)\,]
\left\{\,\sumninf\,a_n\,\left\{\frac{\gamma}{\gamma^2 + \pi^2 n^2}\,\right\} 
+ \frac{a_0 }{\gamma}\,\right\}.\\
&&\label{ou}
\ear

Let's compare the two formulas. 
Look first at (\ref{weiner}), Wiener's formula for generating a process
generally considered to be the epitomy of stochasticity. Yet once the coefficients (the $a_n$)
are chosen, the process is purely deterministic; 
why then can we consider it ``stochastic"? Perhaps because 
the sum in (\ref{weiner}) converges only conditionally and hence is unstable, in some sense.
But then look at (\ref{ou}), our derived Wiener-type formula for the trajectories of the OU process.
Here the sums are absolutely convergent. Yet do we not regard the OU process to be stochastic?
Indeed, some say: {\em the} \BM?

Now let's return to the \qp\ signal of formula (\ref{qpdef}).
First observation: there is no term proportional to `$t$'.
There cannot be, because we have confined everything to the box of fixed width, say, `$A$'.
But this may not be important, since presumably `$A$', the size of the water drop
we are observing, is much larger than the field of the microscope and hence much greater than
observed motions of the pollen grain. 

But is randomness in the initial conditions 
sufficient to justify saying that, over time, the \qp\ formula generates 
 `Brownian-Motion-like'
 trajectories?
Let's speculate about what characteristics of the amplitudes $\{a_n\}$ and the frequencies 
would pass the test:

\begin{quote}
\centerline{\bf Conjecture}

Formula (\ref{qpdef}) will generate `\BML' (BML) trajectories provided:

The $\{a_n\}$ are chosen to be mean-zero, orthogonal, random variables and either:

(a)

\be
\sum\,|a_n| = \infty,
\ee

\ni or:

(b)
\be
\sum\,|a_n|\,|\nu_n| = \infty.
\ee

\ni (or both,
say, almost everywhere, or at least on a set of positive measure).
\end{quote}

The reader may doubt that a \qp\ signal 
can really appear `stochastic'.
Perhaps generating some curves from
formula (\ref{ou}) would be persuasive. In Figure One,
I show five curves generated directly from that formula, with the $a_n$ i.i.d. N(0,1) and
 adding up 1,000 terms of the
sum. 
 Parameter `$\gamma$' was 10.0. 
(The curves are the output of different `runs', 
with different `seeds' to the Random Number Generator, or RNG; 
once the $a_n$ are chosen, the RNG is no longer needed. 
Varying the number of terms in the sum
did not noticeably affect the curves, which is not surprising given the factors of $n^2$
in the denominators.)
Note the many changes of direction
in the curves.

\begin{figure}
\rotatebox{0}{\resizebox{5in}{5in}{\includegraphics{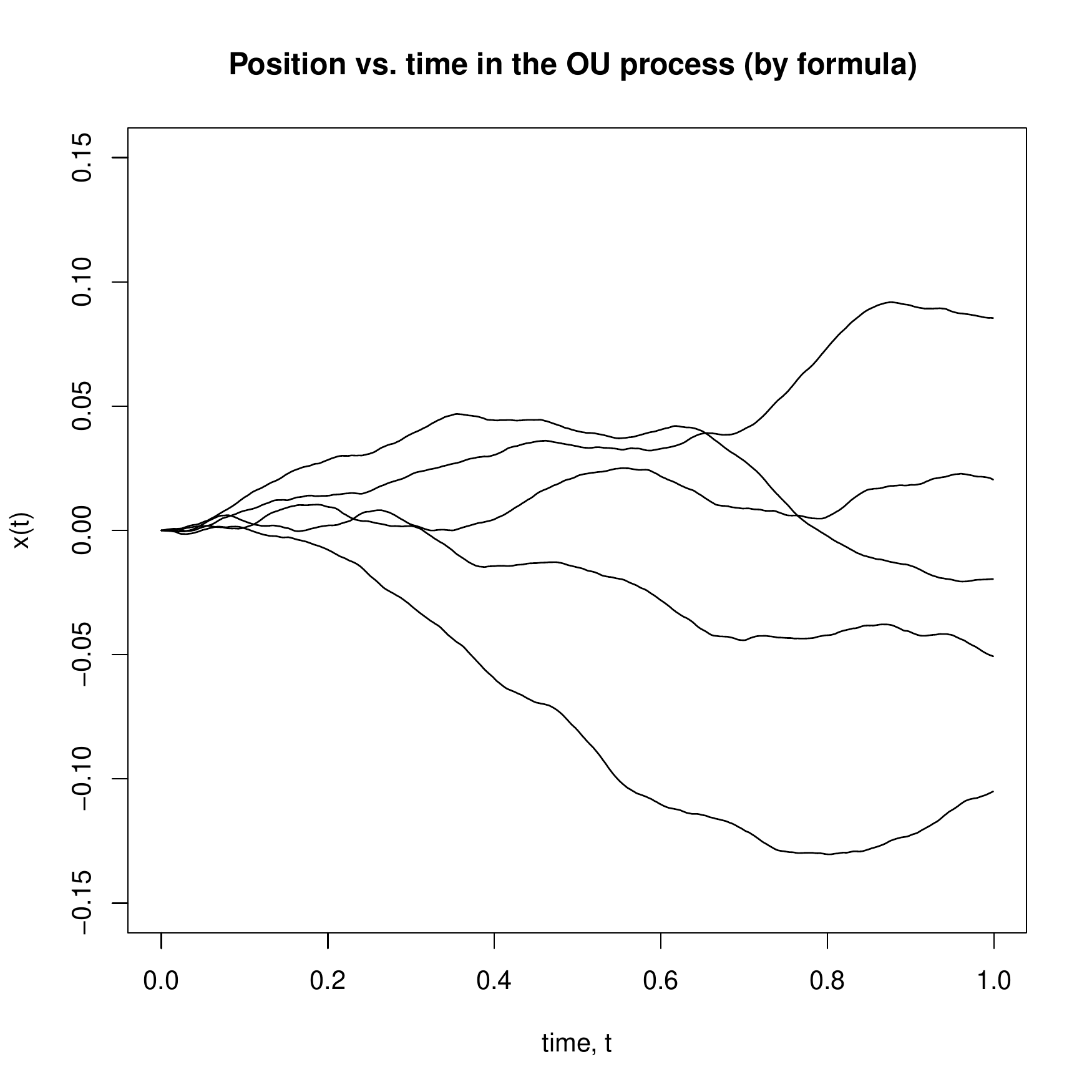}}}
\caption{Position as a function of time, from a formula with i.i.d. normal coefficients.
\label{fig1}}
\end{figure}

Another argument for the positive claim, BML,
is to imagine adding up the series on a computer. For either the
Weiner formula or the time-derivative of the OU formula, this cannot be recommended.
These series are conditionally convergent, meaning that by adding up more 
and more terms the partial sums will
approach some number. But that is due to sign changes, and how many such terms
are necessary for a preselected goal of accuracy might fluctuate wildly with changing time.
Thus examining a graph of output from adding up a thousand terms for each time-point
is likely to be a fool's errand. 

By contrast, in the negative case of the Conjecture (conclusion: 
``not Brownian-motion like", NBML), with absolutely convergent series one can
confidently truncate the sums and run off graphs.
Why NBML? Because a \qp\ signal with a finite number of frequencies should be regarded
as representing an oscillation, albeit perhaps a complicated one.
Curiously, at the time Perrin published his book (1916), \cite{perrin}, 
he had to confront claims by others
that they were all in fact observing an oscillation \cite{perrin}.
Perrin, a supporter of the Poincar{\'e}/Einstein model,
was adamant that this was not the case.
 
I hope the reader is convinced that to say: ``Wavefunctions along with linear
\Schseqn-induced time evolution 
produces \qp\ signals, so can never explain \BM" is a bit glib.

\pagebreak
\subsection{What are the possible conclusions of this research?}

Before introducing models, let's ask for the goals and possible outcomes in this project.
Here we must be wary of epistemological traps.

We might find that, in some model, possibly a toy model in fewer dimensions than three or
with oversimplified potentials, we nevertheless discover \BML\ signals can emerge,
then I would say: game over, \wfs\ and linear dynamics can do the job. 

On the other hand, suppose we fail to discover such signals in our models.
We must be modest about declaring that \wfs\ plus linear dynamics cannot
therefore 
explain \BM. That declaration would be tantamount to a ``no-go" theorem, and the history of these
theorems reveals that they always exhibit circular reasoning 
(covertly assuming the conclusion), as Bell remarked 
in a paper of his in 1966, \cite{Bell66}. Thus such ``proofs" usually
carry no scientific weight (the only exception I would make 
to this rule is for the second version(s)
of Bell's Theorem of 1964, which came along about a decade later). In our case, all we could
conclude is that whatever oversimplified models we considered did not explain the phenomena
(but maybe some more-realistic model might). 

The pessimism contained in the last paragraph might be overcome if we can describe
structural features inevitable in any conventional quantum model of this situation, 
that renders the negative 
conclusion more-or-less ``generic".

\section{Part Two: The Physical Models\label{model_section}}

\subsection{The model class}

For simplicity in writing formulas (avoiding vector notation), 
I will indicate a point in $R^d$ by $x$ or $X$. (As if the space dimension, $d$, is one.
The reader will readily generalize the formulas to three dimensions.)
Let $x_n: n = 1,...,N$ denote the positions of the light particles, and $X$
the position of the heavy particle.
Our wavefunction will have the form:

\be
\psi(x_1,x_2,...,x_N;X),
\ee

\ni and our \Schseqn\ will be:

\bar
\no i\,\hbar\,\frac{\partial \psi}{\partial t} &\=& \left\{\, 
- \frac{\hbar^2}{2m}\,\sumn\,\triangle_n  - 
\frac{\hbar^2}{2M}\,\triangle_X  + \sumn\,v(x_n - X) + \sum_{n\neq m}\,u(x_n-x_m)\,\right\}\,\psi\\
\no &\=& H\,\psi.\\
&& \label{scheqn}
\ear

\ni Here, the $\triangle$ symbols stand for second-derivative operators,
 $v(\cdot)$ denotes a scattering potential, which I take to be positive, symmetric, and tending
to infinity at the origin, and $u(\cdot)$ is any pair potential representing the forces
between the light particles. (It is the latter that would incorporate e.g., viscosity.) 
Also, `$H$' denotes the conventional Hamiltonian (a self-adjoint linear operator).

Naturally, we would like to study the case of $d=3$ and with boundary conditions: 
$\psi = 0$ if any argument ($x_n$ or $X$) lies outside of either a sphere of radius `$A$'
or a disk of that radius and some given thickness (representing the water droplet).

For the initial conditions: we have to get temperature in,
and whatever we insist for initial condition of the heavy particle, discussed in a later section.

Perhaps it would be useful to introduce relative coordinates and the center-of-mass, as
we do for scattering problems?
That will produce a splitting of the Hamiltonian, but introduces a complication.
Anyway, let:

\bar
\no && z_n \= x_n \- X;\\
\no && S \= \hat{m} \sumn\,x_n \+ \hat{M}\,X;\\
\no && \hat{m} \= \frac{m}{Nm + M};\\
\no && \hat{M} \= \frac{M}{Nm + M};\\
&&\label{S_def}
\ear

\ni (Here the hatted masses are called the ``reduced masses" and $S$ is the COM.)
Introducing a new wavefunction with the new arguments by:

\be 
\phi(z_1,...,z_N;S) \= \psi(x_1,...,x_N;X),
\ee

After some work with the chain rule, the Hamiltonian in the new variables becomes:

\bar
\no H\,\psi &\=& \- \frac{\hbar^2}{2\,m}\,\sumn\,\frac{\partial^2\,\phi}{\partial\,z_n^2}\,
\- \frac{\hbar^2}{2\,M}\,\sum_{k,n}\,\frac{\partial^2\,\phi}{\partial\,z_n\,\partial \,z_k}\\
\no && \+ \left\{\, \sumn\,v(z_n) \+ \sum_{k \neq n} u(z_n - z_k)\,\right\}\,\phi \\
\no &&  \- \left\{\,N\,\frac{\hbar^2}{2\,m}\,\hat{m}^2 
\+ \frac{\hbar^2}{2\,M}\,\hat{M}^2\,\right\}\, 
\frac{\partial^2\,\phi}{\partial\,S^2}\\
&&
\ear

So we get the splitting, but the complication is the second term with the mixed partial
derivatives.

Next, what will we observe? Of course, the motion of the heavy particle:

\be
O(t) \= <\psi(t)|X|\psi(t)>.
\ee

Assuming that $H$ has discrete spectrum:

\be
H\psi_k \= \zeta_k\,\psi_k,
\ee

\ni (where $k = 1,2,...$) and letting $\omega_k = \zeta_k/\hbar$, we have, by expanding
 the initial \wf\ using this orthonormal basis as:

\be
\psi(0) \= \sumk\,c_k\,\psi_k,
\ee

\ni then at later times:

\be
\psi(t) \= \sumk\,c_k\,\exp\left\{\,i\,\omega_k\,t\,\right\}\,\psi_n.
\ee

\ni For the observable we have: 

\be
O(t) \= \sum_{j,k}\,c_k\,c_j^{*}\,<\psi_j|X|\psi_k>\,
\exp\left\{\,i (\omega_k - \omega_j)\,t\,\right\}.\label{obs}
\ee

Using the relative coordinates and the splitting, we can write:

\be
H \= H_z \+ H_S,
\ee

\ni for the two parts involving the $z_n$ and  $S$ respectively.
However, we do not get a splitting of the eigenvalues as a sum of those associated
with each operator. The reason is that the boundary conditions do not transform into
a square or rectangle in the relative coordinates, but rather 
a tilted figure. (E.g., let $N=1$ and boundary conditions:
$-A < x_1 < A$ and $-A < X < A$. The four vertices of the square of side $A$ map into the
points in the $(S,z_1)$ plane: $(A,0)$, $(-A,0)$, $(\left\{\,\frac{m-M}{m+M}\,\right\}\,A,2A)$
and $(\left\{\,\frac{M-m}{m+M}\,\right\}\,A,-2A)$; because the transformation is linear,
we obtain the bounded region with these points as vertices, hence a tilted parallel piped.)

We can, however, use conservation of total momentum to express the observable in terms
of the light particles, if desired. Noting that

\be
X \= \left(\,\frac{1}{M}\,\right)\,\left\{\,(Nm+M)\,S \- m\,\sumn\,x_n\,\right\},
\ee

\ni and assuming that $<\psi(t)|S|\psi(t)> = 0$ 
(which we can always 
do by selecting a suitable reference frame in which to view the system) we have:

\be
O(t) \= - \left(\,\frac{m}{M}\,\right)\,<\psi(t)|\sumn\,x_i|\psi(t)>.
\ee

\subsection{A toy model in one spatial dimension\label{1dtm}}

Call it the ``One-Dimensional Toy Model" (1DTM).
The idea is to avoid potentials by imagining an infinitely-high wall preventing the
light particles from tunneling past the heavy particle. Our \wf\ now takes the form:
 
\be
\psi(x_1^R,x_2^R,...,x_N^R;x_1^L,x_2^L...x_N^L;X),
\ee

\ni where the $x_n^{R,L}$ are the positions of the light particles to the
right or left of the heavy particle, respectively.

The \Sch's equation takes the form:

\be
i\,\hbar\,\frac{\partial \psi}{\partial t} \= \,\left\{\, 
- \frac{\hbar^2}{2m}\,\sumn\,\left\{\,\triangle_n^R + \triangle_n^L\,\right\} - 
\frac{\hbar^2}{2M}\,\triangle_X  \right\}\,\psi.\label{scheqn1}
\ee

For boundary conditions we take:

\be
-A < x_n^R < A; \ph -A < x_n^L < A;\ph -A < X < A;\ph x_n^R > X;\ph x_n^L < X.
\ee

\ni (meaning that the \wf\ is zero if any of the conditions are violated). 

It was my hope that this model would be exactly solvable. But there was a curious obstacle
to finding formulas for the eigenvalues. This concerns the old question, which Marc Kac
advertised in 1966, \cite{Kac}: ``Can you hear the shape of a drum?" 

Consider just one light particle, say to the right of the heavy particle. The boundary conditions
in the $(X,x_1)$ plane then yield a triangle with vertices at $(-A,-A)$, $(-A,A)$ and $(A,A)$.
By rescaling the variables, defining: 

\be
y_1 = \sqrt{\frac{2m}{\hbar^2}}\,x_1; \ph Y = \sqrt{\frac{2M}{\hbar^2}}\,X,
\ee

\ni then writing: $\psi(x_1,X) = \phi(y_1,Y)$, if the latter satisfies the Laplacian
eigenvalue equation:

\be
\frac{\partial^2 \phi}{\partial y_1^2} \+ \frac{\partial^2 \phi}{\partial Y^2} = \lambda \phi,
\ee

\ni then $\psi$ will satisfy our \Schist\ eigenvalue problem. 
For boundary conditions on $\phi$ we get another
triangle, stretched in both axes.    

Thus, we are asking about whether we can hear the shape of a drum in the form of a right triangle.
But here I hit the obstacle: apparently, this problem is unsolved! The only cases I could
find in the literature yield formulas for the equilateral triangle, the hemi-equilateral triangle
(30-60-90), and the isoceles right triangle (45,45,90), but none for the general right triangle,
\cite{KandS,BJC}. 

Hence, short of solving a 100-year-old open problem, we cannot expect to possess formulas
for the eigenvalues and eigenfunctions. However, we can get some upper and lower bounds
on the former by exploiting the Domain Monotonicity Property of the Dirichlet problem:
let there be two regions, $R$ and $R'$, in the space of 
the positional variables with $R \subset R'$.
Let $\zeta_1 < \zeta_2 < \dots$ and $\zeta_1' < \zeta_2' < \dots$ 
be the corresponding eigenvalues. Then:

\be
\zeta'_1 < \zeta_1; \ph \zeta'_2 < \zeta_2, \dots .
\ee

\ni (Mnemonic: ``The smaller drum has the higher frequencies.") 

To apply the DMP we need an immersed rectangular and a surrounding rectangular region.
For the immersed we replace the b.c.'s by choosing two numbers $B$ and $C$
with $-A < B < C < A$ 
and adopt the b.c.'s:

\be
 -A < x_j^L < B;\ph C < x_k^R < A; \ph B < X < C.
\ee

\ni We can then form eigenfunctions for this region by, for any collections of integers
of form $(n_1^R,...,n_N^R;n_1^L,...,n_N^L;n^X)$:

\be
\prod_{j,k}^N\,\sin\left(\,\pi\,n_j^L\,\left[\,\frac{x_j^L + A}{B+A}\,\right]\,\right)  
\,\sin\left(\,\pi\,n_k^R\,\left[\,\frac{x_j^R -C}{A-C}\,\right]\,\right)  
\,\sin\left(\,\pi\,n_m^X\,\left[\,\frac{X - B}{C-B}\,\right]\,\right)  
\ee

The corresponding eigenvalues are:

\be
\frac{(\hbar\pi)^2}{2m}\,\left\{\,\left(\,\frac{1}{B+A}\,\right)^2\,\sum_{j=1}^N\,(n_j^R)^2 +
\left(\,\frac{1}{A-C}\,\right)^2\,\sum_{k=1}^N\,(n_k^L)^2\,\right\} +
\frac{(\hbar\pi)^2}{2M}\,\left(\,\frac{1}{C-B}\,\right)^2\,(n^X)^2.
\ee

\ni These expressions are upper bounds on the eigenvalues 
of the problem with the triangular region. 
Lower bounds can be derived from the surrounding rectangle:

\be
 -A < x_j^L < A;\ph -A < x_k^R < A; \ph -A < X < A.
\ee

The corresponding eigenvalues are:

\be
\left[ \,\frac{(\hbar\pi)^2}{2m}\,\left\{\,\sum_{j=1}^N\,(n_j^R)^2 +
\sum_{k=1}^N\,(n_k^L)^2\,\right\} +
\frac{(\hbar\pi)^2}{2M}\,(n^X)^2\,\right]\,\left(\,\frac{1}{2A}\,\right)^2.
\ee

If we choose three equal intervals:

\be
B+A \= A - C \= C - B,
\ee

We find the upper bounds are 9 times the lower bound.

\section{\Schist\ {\em vs.} \Copist\ Theory of Observation\label{obs_section}}

The reader may have been surprised by the claim that what we observe of the pollen grain is:

\be
x(t) = <\psi(t)|X|\psi(t)>.\label{observePG}
\ee

In the Copenhagenist philosophy (found in most textbooks), this quantity is called ``the
expected value of the position (of the grain)". The doctrine claims that $\psi$ is a peculiar
(complex-valued!) statistical quantity and that ``the probability of `finding' the grain
in a (spatial) domain $D$ at time `$t$' 
is given by"\footnote{The substitution of `finding' for `being'
or `existing' reflects a positivist ambivalence about whether the particle actually
has a location, if no one is looking.}:

\be
<\psi(t)|P_D|\psi(t)>
\ee

\ni where $P_D$ denotes the projection operator onto the domain, here simply multiplication
by the indicator function: $1[X \in D]$. 

One serious objection to the \Copist\ interpretation is the problem of the timing of observations.
Suppose we look into the microscope at a time $\delta t$ after noon ($t=0$), 
when we prepared the experiment. Suppose we arranged
the initial state, call it $\psi_0$, so that ``the grain lies in $D$ with certainty", 
meaning $P_D\psi_0 = \psi_0$. Then, at the later time, the probability that the grain is still
'found' in the domain is:

\bar
\no <\psi(t)|P_D|\psi(t)> &\=&
 <\exp\{iH\delta t\}\psi_0|P_D|\exp\{iH\delta t\}\psi_0>\\
\no &\=& <\psi_0|\exp\{-iH\delta t\}P_D\exp\{iH\delta t\}\psi_0>\\
\no &\=& <\psi_0|P_D\exp\{-iH\delta t\}P_D\exp\{iH\delta t\}\psi_0>\\
&&\label{probD}
\ear

\ni where I have used that $P_D$ is self-adjoint and $P_D\psi_0 = \psi_0$.
\Copists\ also assert that, having `found' the grain in the domain,
the wavefunction is (must be?) collapsed to a wavepacket supported entirely in that domain.
 
Now here comes the problem: if we expand the exponentials in (\ref{probD}) like:

\be
\exp\{iH\delta t\} = I + iH\delta t - (1/2)\left(\,H\delta t\,\right)^2 + \dots,
\ee

\ni there is no nonvanishing term in (\ref{probD}) proportional to $\delta t$! 
Hence, it is impossible to obtain here an exponential-decay law,
and worse, if we take $\delta t$ small and repeat the observations at successive intervals,
we can freeze the grain in its original domain! (This strange prediction
of \Copist\ quantum mechanics is called 
the ``quantum Zeno paradox" or sometimes 
 the ``watched-pot never boils paradox", or does it really exist and so we should
sustitute ``effect" for ``paradox"? For the history of disputing this claim, see the Appendix.)
 
One way out of this dilemma is to remark that observations cannot be continuous,
because, as neurobiologists have shown, a small but definite time is rquired to acquire
any {\em Gestalt}. But then how can we make predictions about the grain's movement from our 
theory, short of articulating a detailed model of the observer's eye and brain? 
Here is a cogent, pragmatic reason to reject the \Copist\ doctrine of measurement.

Here is another: even if physiology provides us with 
an answer to the question: ``how frequent" (are collapses), 
what about the ``how small" issue, refering to the size of domain $D$? 
I suppose you might take
$D$ to represent the smallest spatial scale your eye+microscope can resolve. But
now I acquire a new microscope of twice the resolving power. 
Must I now shrink the domain by half? 
Each such replacement will further ``chill" the grain, decreasing its movement.
Clearly, in our context of continuous variables and observations, 
\Copism\ is too ambiguous to guide us. 
 
\Schists\ flatly reject the statistical interpretation of the wavefunction.
For us, a single wavefunction is a configuration of matter, 
with no connection whatever to probability or statistics (except if we adopt an ensemble of such), 
and a measurement is described by some functional of it, as for $x(t)$ above. 
Of course, other choices are conceivable, even (\ref{observePG}) 
provided it can be linked to 
the registering apparatus. Even then, we would not label it a ``probability"; not every
real number between zero and one deserves that interpretation. Nor would we
interpret an expression like

\be
<\psi(t)|P_{D'}\,H\,P_D|\psi(t)>
\ee

\ni (where $D'$ is another domain) as ``the transition rate" or ``rate of jumping" 
from $D$ to $D'$. In no formulation known to me is Quantum Mechanics a theory of a 
stochastic jump process. 
\Schists\ do not believe in ``collapses" or ``projections" or ``quantum jumps"; 
we believe in wavefunctions and entanglements.

\section{Initial Conditions and the Thermal Ensemble: Several Choices\label{initial_section}}

Concerning the initial conditions, two scenarios can be imagined: call them the
``Equilibrium Scenario" (EqS) and the ``Non-Equilibrium Scenario" (NEqS).
In the EqS, we postulate that the whole system, water drop plus pollen grain,
is initially in thermal equilibrium at some temperature. 
At zero time, we aim our microscope at the droplet,
notice the pollen grain, and begin to observe it (without making any significant perturbation 
of the system).\footnote{I think even \Copists\ would agree with this, as the pollen
grain is nearly macroscopic and so ``quantum back-action" should be minimal.}
In the NEqS, the water drop is initially in thermal equilibrium; 
at time 0 we drop in a pollen grain and then follow it afterwards.

Next, there is the choice of ensembles, \Copist\ 
or \Schist, with different implications for each scenario.
Consider first the EqS. The \Copist\ predeliction\footnote{ 
Yielding what historically was called,
even prior to the revolution of 1924-6, 
``quantum statistics".} 
is that a thermodynamic system should 
be (would be found?)
in an eigenstate, say with energy $\zeta_n$, weighted in the ensemble by the Gibbs factor:

\be
\exp\left\{\,- \zeta_n/kT\,\right\}.
\ee

\ni where `$k$' denotes Boltzmann's constant and `$T$' the temperature. 
But this makes no sense for the \BM\ problem in the EqS, 
as eigenstates are stationary states so nothing would happen. 

\Schists\ will naturally choose a Gibbsian thermodynamic ensemble of wavefunctions, 
as in
De Carlo and Wick \cite{decarlowick} (who treated a discrete-spin scenario). 
This may be formally
expressed by (identifying as usual each wavefunction with a list of coefficients, e.g.,
$\psi = \sum\,c_k\,\psi_k$), for any bounded functional of the wavefunction, `$f$': 

\bar
\no \cE_T\,\left[\,f(c_1,c_2,...,)\,\right] &\=& Z^{-1}\,\int_{\{\sum\,|c_k|^2 = 1\}}\,\prod\,dc_k\,\exp\left\{\,-
\beta\,\sum\,|c_k|^2\,\zeta_k\,\right\}\,f(c_1,...)\\
\no Z &\=& \int_{\{\sum\,|c_k|^2 = 1\}}\,\prod\,dc_k\,\exp\left\{\,-
\beta\,\sum\,|c_k|^2\,\zeta_k\,\right\}.\\
&&
\ear

\ni (Here $\beta = 1/kT$; 
so $\cE_T$ stands for expectation over wavefunctions, in a Gibbs ensemble with temperature
`$T$'.)

Because we are in infinitely-many dimensions, it would be best to restrict the
integrals to a subspace of maximal allowed  
energy\footnote{The 
mathematical issue here is that the unit ball of an infinite-dimensional 
Hilbert space is not compact,
rendering problematic integration over it. With the energy bound,
provided $\zeta_n \to \infty$, the set is compact. },
e.g., to:

\be
 \sum\,|c_k|^2 =1;\ph \sum\,|c_k|^2\,\zeta_k < E_{\hbox{max.}}.\label{energycondition}
\ee

In the NEqS, it would be natural to choose the initial condition as approximately:

\be
\psi_0 \= \phi(x_1,x_2,....,x_N)\cdot\delta_0(X),
\ee

\ni where $\phi(\,)$ might be chosen by a \Schist\ from the Gibb ensemble at temperature `$T$'
with energy function (Hamiltonian) lacking the terms in derivates with respect to `$X$',
while a \Copist\ would presumably assume that $\phi(\,)$ is an eigenstate of that
abbreviated Hamiltonian, with weight in the ensemble given the the corresponding Gibbs factor.

I will only consider the EqS in this paper, reasoning that Brown and Perrin probably
squeezed a droplet onto a microscope slide from a bottle containing water and pollen 
already at equilibrium (rather than sprinkle pollen over the wetted slide).

\section{BML or NBML?\label{theorem_section}}

We have defined our observable to be:

\bar
\no O(t) &\=& <\psi(t)|X|\psi(t)>\\
\no &\=& \sum_{j,k}\,c_k\,c_j^{*}\,<\psi_j|X|\psi_k>\,
\exp\left\{\,i (\omega_k - \omega_j)\,t\,\right\}\\
\no &\=& <\psi(0)|X|\psi(0)> + 
 \sum_{j \neq k}\,c_k\,c_j^{*}\,<\psi_j|X|\psi_k>\,
\exp\left\{\,i (\omega_k - \omega_j)\,t\,\right\}.\\
&&
\ear

\ni We then relabeled the second (non-constant) term as:

\be
\sum_n\,a_n\,\exp\left\{\,i \nu_n\,t\,\right\}.
\ee

We assume the boundary conditions in the models described in section \ref{model_section},
the normalization of the $\{c_k\}$, and the energy bound stated in section \ref{initial_section}.
Also, for simplicity, one dimension (there is no difficulty in extending to three dimensions).
For one factor in the expression above we can derive another expression (see Math Appendix):

\be
<\psi_j|X|\psi_k> \= \left(\,\frac{1}{\zeta_j - \zeta_k}\,\right)\,\frac{\hbar^2}{2M}\,
\left\{\,<\frac{\partial \psi_j}{\partial X}|\psi_k> -
<\psi_j|\frac{\partial \psi_k}{\partial X}> \,\right\}.\label{expression}
\ee

\ni Define:

\be
B_{j,k} \= \frac{\hbar^2}{2M}\,
\left\{\,<\frac{\partial \psi_j}{\partial X}|\psi_k> -
<\psi_j|\frac{\partial \psi_k}{\partial X}> \,\right\}.
\ee

We can now state the following theorem:

\begin{quote}
{\bf Theorem} The observed trajectories will be NBML if:

\be
\sum_{j,k}\,\frac{|B_{j,k}|^2}{\zeta_j\,\zeta_k}\ph < \ph \infty.\label{theorem}
\ee

{\bf Corollary} The above conclusion follows if:

\be
\sum_j\,\frac{1}{\zeta_j} \ph< \ph\infty.\label{corollary}
\ee
\end{quote}

By `NBML' we mean that both sums in the Conjecture of section \ref{formula_sect} are finite.
For the proof of the Theorem and Corollary, see the Math Appendix.

In one model introduced in the paper, namely the 1DTM, the
assumption (\ref{corollary}) of the Corollary probably doesn't hold. See the Math Appendix
for the reasoning. 
The intuition supporting the theorem as stated is that the eigenfunctions have different spatial
frequencies, and so there may be additional fall-off in the coefficients $B_{j,k}$ for
$|j-k|$ large.

\section{Diffusive or Ballistic?\label{diffusive_section}}

The BML {\em vs.} NBML distinction referred only to apparent randomness of trajectoires. What about
average displacement? 

In our set-up we are observing

\bar
\no x(t) &\=& \sum_{j,k}\,c_k^*\,c_j\,<\psi_k|X|\psi_j>\,\exp\{\,i(\omega_j-\omega_k)\,t\,\}\\
\no  &\=& \sum_{j,k}\,c_k^*\,c_j\,g_{k,j}\,\exp\{\,i(\omega_j-\omega_k)\,t\,\}\\
&&
\ear

\ni where I have intoduced an abbreviation for the matrix of integrals. 
For the mean-square displacement (MSD) we have:

\bar
\no \cE\,|x(t) - x(0)|^2 &\=& \sum_{j,k,r,s}\,\cE\left\{\,c_k^*\,c_j\,c_r^*\,c_s\,\right\}\,
g_{k,j}\,g_{s,r}^*\,\left[\,\exp\left\{\, i(\omega_j - \omega_k + \omega_s - \omega_r\,)\,t
\right\}\right.\\
\no &&\left. - \exp\left\{ i(\omega_j - \omega_k)\,t\,\right\}
- \exp\left\{-i(\omega_r - \omega_s)\,t\,\right\} \+ 1\,\right].\\
&&\label{wehave}
\ear

\ni Here $\cE$ denotes expectation over an ensemble of initial conditions. 

Consider first the EqS and the \Schist\ Gibbs ensemble.
In this case, any expression of form:

\be
\cE\left\{\,c_j^*\,c_k\,c_r^*\,c_s\,\right\}
\ee

\ni vanishes if any coefficient is unpaired with another (meaning different indicies).
For our Gibbs ensemble this follows by the rotational invariance of
the measure on the sphere and that the energy only depends on the $\{|c_k|^2\}$ (since
the mapping: for any given $j$,
$c_j \longrightarrow \exp\{i\alpha\}\,c_j$ with real $\alpha$ 
and the other coefficients unchanged
preserves both the measure and the energy). Same if the pairing is not by a coefficient and its
complex congugate.

Searching for coefficient coincidences in (\ref{wehave}), 
the cases: $k=j$, $r=s$ and $j=k=r=s$ yield zero,
and the only non-zero contribution is from $k=s$, $j=r$, yielding:

\be
\cE\,|x(t) - x(0)|^2 \= 2\,\sum_{j,k}\,\cE\,\left(\,|c_k|^2\,|c_j|^2\,\right)\,|g_{k,j}|^2
\,\left\{\,1 - \cos[(\omega_j - \omega_k)t]\,\right\}.\label{result}
\ee

In the NEqS, we might postulate:

\be
\psi_0 \= \phi(x_1,x_2,...,X_N)\,\theta(X),
\ee

\ni with $\theta(\cdot)$ very narrow near the origin. Then for later times:

\bar
\no \psi(t) &\=& \sum_k\,c_k\,\psi_k\,\exp\left\{\,i\,\omega_k\,t\,\right\};\\
\no c_k &\=& <\psi_j|\psi_0>.\\
&&
\ear

\ni where as usual the $\psi_k$ and $\omega_k = \zeta_k/\hbar$ are the eigenfunctions and
corresponding frequencies and energies. Writing

\be
\phi \= \sum_j\,g_j\,\phi_j
\ee

\ni where the $\phi_j$ are the eigenfunctions of the water molecules dynamics absent
the pollen grain, we have:

\be
c_k \= \sum_j\,g_j\,<\psi_k|\phi_j\,\theta>.
\ee

If we adopt a Gibbs thermal ensemble for the water before we drop in the grain:

\be
\cE_T\,f(c_1,c_2,...,c_N) \= Z^{-1}\,\int\,\prod\,dg_j\,\exp\left\{\,- \beta\,\sum |g_j|^2\,\eta_j
\,\right\}\,f(c_1,...),
\ee

\ni (where the $\eta_j$ are the eigenvalues absent the grain) then $\cE_T c_k = 0$ 
as before, but now:

\be
\cE_T\,c_1\,c_2^* \= Z^{-1}\,\int\,\prod\,dg_j\,\exp\left\{\,- \beta\,\sum |g_j|^2\,\eta_j
\,\right\}\,\left[\,\sum_{j,s}\,g_j\,g_s^*\,<\psi_k|\phi_j\,\theta>\,<\psi_k|\phi_s\,\theta>^*
\,\right],
\ee

\ni there is a non-zero contribution from the sum in square brackets for $j=s$.
So we will not get the simple result in (\ref{result}).

\ni Now we can ask: in either of these scenarios, 
can the MSD of the pollen grain behave like `$t$'?
We already know that it is impossible for long times, because everything is confined to a box.
Perhaps we could have that behavior for some initial
time period, because, expanding in a Taylor's series
in `$t$', the term of second order has a small coefficient,
 relative to the first? But the answer is no also,
because, expanding the square-bracketed terms in (\ref{wehave}) in such a series, 
the first non-constant term
contains $t^2$. This conclusion is not dependent on the initial condition, 
or choice of ensemble for such.

But now recall the situation for the OU process near $t=0$, 
given by expanding the exponentials in (\ref{OUMSdisplacement}). 
The term proportional to `$t$' is absent; thus, the diffusive behavior
in the OU process exists only for long times.
But for our models, the motion of
the grain (and the water molecules) is limited to the water droplet, 
so the MSD cannot grow like `$t$' forever. 
Nevertheless, it remains a possibility that there is an intermediate time scale on which diffusion
is observed in our models.

\section{WFE, Blocking `Cats', and All That\label{WFEsection}}

Within the \Schist\ paradigm,
it might be argued that the reason for negative conclusions about pollen-grain trajectories
(i.e., NBML)  
is that the assumed ensemble is made up predominantly
by ``cat states": wavefunctions in which the dispersion of the center of mass of 
the system is larger than its size. Perhaps BML trajectories would appear
if the wavefunctions could be restricted to be narrow in configuration space, i.e.,
be more ``classical"?
We would need some mechanism to prohibit cat states. 
In \cite{WickI}, the author suggested one that may be called ``WaveFunction Energy" (WFE),
which is of form

\def\cHWFE{\cH_{\hbox{WFE}}}
\def\cHQM{\cH_{\hbox{QM}}}
\be
\cHWFE \= w\,N^2\,D(\psi),
\ee

\ni where $D(\psi)$ is the dispersion of the center-of-mass of the system:

\be
D(\psi) \= <\psi|S^2|\psi> - <\psi|S|\psi>^2,
\ee

\ni Here, $w$ is a positive constant and 
$S$ is given by (\ref{S_def}). The energy functional\footnote{What the 
Copenhagenists call ``the expected energy", but just ``the energy" for \Schists.}
is then given by:

\be
\cH \= \cHQM \+ \cHWFE,
\ee

\ni and the corresponding \Schseqn\ is given by:

\bar
\no i \hbar\,\frac{\partial \psi}{\partial t} &\=& \frac{\partial}{\partial \psi^*}\,\cH\\
\no &\=& H_{\hbox{QM}}\,\psi \+ w\,N^2\,\left\{\,S^2\,\psi \- 2\,<\psi|S|\psi>\,S\,\psi\,\right\}.\\
&&
\ear

This \Schseqn\ is nonlinear, and so the eigenfunction-eigenvalue analysis made for the
linear models is not available. In \cite{WickIII}, the author noted that, in a measurement
scenario involving a double-well external potential acting on the observed part, there
is an instability not present in any linear model which suggests the kind of sensitive
dependence on initial conditions seen in so-called ``chaotic" models (or for that matter
in the betting game of roulette). Thus it is possible that the dynamics generated by
the model with WFE might exhibit more random-looking orbits than in linear models.

What is the effect of nonlinear dynamics 
on the conclusion of no-diffusive behavior? We can't here make the
kind of calculation of section \ref{diffusive_section}, but there is another approach. Consider the
quantity:

\be
x(t) - x(0) \= <\psi(t)|X|\psi(t)> - <\psi(0)|X|\psi(0)>.\label{quantity}
\ee

If we are to observe diffusive behavior at least for small times, 
it is necessary that this quantity have
an expansion in terms of $\sqrt{t}$, i.e.,

\be
x(t) - x(0) \= a\,\sqrt{t} \+ b\,t \+ c t^{3/2} \+ \cdots
\ee

which in turn would imply:

\be
\frac{\partial}{\partial t}\Big|_0\,\left[\,x(t) - x(0)\,\right] = \infty.\label{inf_eqn}
\ee

But this does not hold. In fact,

\be
\frac{\partial}{\partial t}\Big|_0\,\left[\,x(t) - x(0)\,\right] \= \frac{i}{\hbar}\,
<\psi(0)|\,\left[\,\cH,X\,\right]\,|\psi(0)>,\label{commutator}
\ee

\ni where here $[\cdot]$ denotes the commutator. 
The new terms in the nonlinear \Schseqn\ commute with $X$ and so do not contribute,
while the usual Hamiltonian operator commutation with it yields an expression with has an energy
bound given in (\ref{Ebound}) of the Math Appendix. 
 
Replacing $S$ by $X$, i.e, to restrict only the dispersion of the observed part, 
makes no difference. In some papers, especially when considering relativity 
(both special, in \cite{WickI}, or general,
in \cite{WickV}), the author has considered a momentum form of WFE, obtained by replacing $S$ 
by $P$, the total momentum:

\be
P \= \frac{\hbar}{i}\,\left\{\,m\,\sum_k\,\frac{\partial}{\partial x_k} \+ M\,\frac{\partial}{\partial
\,X}\,\right\}.
\ee

\ni For the commutator we find:

\be
[\,P^2 - 2\,<\psi|P|\psi>\,P,\,X\,] \= 2\,M\,\left\{\,P - <\psi|P|\psi>\,\right\},
\ee

\ni which, plugged into (\ref{commutator}), reveals this choice of nonlinearity 
does not contribute to the derivative at $t=0$ either,
leaving the conclusion unchanged.

Even if WFE does not appear directly in the formulas for 
position or velocity of the grain, it will
affect other components which will feed back onto the former.

Another important
observation is on how we interpret the observable:

\be
O(t) \= <\psi(t)|X|\psi(t)>.
\ee

Suppose we see at some time $O(t) \approx 0$. Without WFE (or some other hypothesis
diminshing the participation of `cat' states), two scenarios might occur.
It might be that the wavefunction is concentrated around the origin (more precisely,
the reduced density function for the variable $X$ is so concentrated); perhaps:

\be
<\psi(t)|P_D|\psi(t)>  \ph \approx  \ph 1,
\ee

\ni where $D$ is some small domain containing the origin. Moreover, if $D'$ and $D''$
are two domains disjoint from $D$ and each other and well-separated:

\be
<\psi(t)|P_{D'}|\psi(t)>  \ph \approx 0  \ph \ph \hbox{and} \ph
<\psi(t)|P_{D''}|\psi(t)>  \ph \approx  \ph 0.
\ee

But without cat-blocking it is consistent with our hypothesis $O(t) \approx 0$
that:

\be
<\psi(t)|P_{D'}|\psi(t)>  \ph \approx  \ph \half \ph \hbox{and} \ph
<\psi(t)|P_{D''}|\psi(t)>  \ph \approx  \ph \half.\label{notCop}
\ee

Suppose that by eye we could distinguish the grain location on the scale or these sets;
then by moving our eye around we could notice that the grain appears 
at two places at once.\footnote{(\ref{notCop}) should {\em not} 
be interpreted as saying: ``The grain appears in either $D'$
or $D''$ with probabilities: 1/2-1/2"; that's Copenhagenism, not \Schism.}
We should be very surprised. Of course, that doesn't happen (nor are cats ever
simultaneously alive and dead). That is why we could interpret $O(t)\approx 0$ as 
meaning, ``The grain is near the origin." WFE (producing an energy barrier to cat formation)
is one mechanism assuring us of this `classical' behavior of pollen grains, independently
of whether they exhibit \BM\ or not.

\section{Discussion}

Concerning the issue of BML {\em vs.} NBML trajectories, as I have stated the question
in terms of series of sines and cosines and their convergence properties, one objection
runs as follows. Even if both series in the Conjecture converge absolutely, allowing one
to truncate them in computations uniformly over any time period, nevertheless 
if the number of terms and corresponding frequencies is large the signal might still
resemble that of a stochastic process, i.e., be BML. I cannot refute this suggestion,
except to note that I know no criterion to differentiate a \qp\ time signal from the latter
based on the number of frequencies.

I have shown, for a few models, that BML-trajectories are possible, even likely. (Unfortunately
I cannot do better, as I cannot compute, e.g., the characteristic frequencies.) 
But diffusive behavior
seems to be  
problematic. There is one scenario still 
conceivable. We know that the mean-squared
displacement must grow quadratically near $t=0$, and since the system is bounded must cease
to grow at long times. 
But, as mentioned in section \ref{diffusive_section}, there might exist an interval of time 
in which MSD growth is approximately linear. Locating that interval in specific models will be
difficult, and even if possible, would it be a convincing demonstration of \BM?
Or would this scenario be interpreted as ballistic motion up until a time when something 
(friction? edge effects?)
slowed it to a crawl?

The inability to derive a diffusion constant of course spoils 
the Einstein program to link the diffusion constant to the viscosity and the
temperature.
To make progress, we really need exactly-solvable models.

\section*{Appendix: Is it a ``Quantum Zeno Effect" or a ``Quantum Zeno Paradox"?}

It has been claimed that this peculiar scenario has been observed in an experiment;
see \cite{MandS}, and, for the history of responses, \cite{Itano} and references therein. 
Hence it must be an ``effect", 
as no one can {\em observe} a paradox. But the situation is difficult
to interpret, especially for a \Schist\ such as the author.

The experimental system was a multilevel atom. A transition between levels one and three
could be detected as laser-induced fluoresence. A long RF pulse could drive the
atoms into a state two, eliminating the signal.  However, during that time, 
short pulses at another frequency could drive transitions between states one and two.
The latter are regarded as ``peeks" and as their number goes to infinity the atoms are
frozen in state two. 
The experimenters observed a reduction in fluoresence with the number of said pulses.

A lot of issues arising here. First, the ``photons" emitted during the short pulses
were not detected; nevertheless, the theorists imposed a projection (``collapse") after each pulse.
The authors replied to this criticism by remarking that they {\em might} have been detected,
justifying the projection postulate. In the context of this paper: perhaps,
if we never pointed our microscope at the droplet and peeked, but nevertheless {\em might have},
should the wavefunction be collapsed? If someone {\em might have} watched an arrow in flight,
it cannot move? Here's a paradox that would have made Zeno proud!

Second, we \Schists\ do not, cannot, use the language of ``populations of atoms in various states",
because we believe only in the wavefunction. Populations and random jumps properly belong
to a mathematical discipline called ``stochastic jump-processes" (useful, e.g., in biology), 
where evidently quantum mechanics does not fit. 

Finally, to precisely interpret this experiment
we would need a comprehensive theory of atoms, the electromagnetic field, and the 
measuring apparatus (in a combined wavefunction, of course). 
Perhaps the correct understanding of this experiment is that such a theory will
have interesting nonlinearities which can be approximated with projection operators,
and the only real observation (of laser-induced fluoresence) was made at the end.

\section*{Math Appendix}

{\bf Proofs of the Theorem and Corollary of section \ref{theorem_section}.}

We first prove (\ref{expression}):

\bar
\no \zeta_k\,<\psi_j|X|\psi_k> &\=& <\psi_i|X|H\,\psi_k>\\
\no &\=& <\psi_j|X|\left\{\, - \sum_n\,\frac{\hbar^2}{2m}\,\frac{\partial^2}{\partial x_n^2}
- \frac{\hbar^2}{2M}\,\frac{\partial^2}{\partial X^2}+V\,\right\}\,\psi_k>\\
\no &\=& + \frac{\hbar^2}{2M}\,<\psi_j|\frac{\partial \psi_k}{\partial X}>
- \frac{\hbar^2}{2M}\,<\frac{\partial \psi_j}{\partial X}|\psi_k> + <H\,\psi_j|X|\psi_k>\\
\no &\=& + \frac{\hbar^2}{2M}\,<\psi_j|\frac{\partial \psi_k}{\partial X}>
- \frac{\hbar^2}{2M}\,<\frac{\partial \psi_j}{\partial X}|\psi_k> + \zeta_j<\,\psi_j|X|\psi_k>\\
&&
\ear

\ni (where `$V$' stands for all the terms involving potentials, and 
we have integrated-by-parts and used the boundary conditions several times) 
from which the claim follows.

We can write:

\be
a_n \= \sum_{j \neq k}\,c_k\,c_j^*\,<\psi_j|X|\psi_k>\,1[\,\omega_k - \omega_j = \nu_n\,],
\ee

\ni (where $1[\cdot]$ denotes the indicator function of the set or condition). Then using
(\ref{expression}):

\be
\hbar\,a_n\,\nu_n \= \sum_{j \neq k}\,c_k\,c_j^*\,B_{j,k}\,1[\,\omega_k - \omega_j = \nu_n\,],
\ee

We can derive the inequalities:

\bar
\no \hbar\,\sum_n\, |a_n|\,|\nu_n| &\=& \sum_n \,|\, \sum_{j \neq k}\,c_k\,c_j^*
\,B_{j,k}\,1[\,\omega_k - \omega_j = \nu_n\,]\,|\\
\no &\leq& \sum_n \, \sum_{j \neq k}\,|c_k|\,|c_j^*|
\,|B_{j,k}|\,1[\,\omega_k - \omega_j = \nu_n\,]\\
\no &\=& \sum_{j \neq k}\,|c_k|\,|c_j^*|
\,|B_{j,k}|\\
\ear

Now the claimed inequality (\ref{theorem}) follows by applying Cauchy-Swartz twice, first to the
sum over `$k$', then over `$j$':

\bar
\no \sum_{j \neq k}\,|c_k|\,|c_j^*|
\,|B_{j,k}| &\leq& 
 \sum_k\,|c_k|\,\sqrt{\zeta_k}\,\frac{1}{\sqrt{\zeta_k}}\,\sum_j\,|c_j^*|
\,|B_{j,k}| \\
\no &\leq& \left(\,\sum_k |c_k|^2\,\zeta_k\,\right)^{1/2}\,\left\{\,\sum_k \frac{1}{\zeta_k}
\left(\,\sum_j\,|c_j|\,|B_{j,k}|\,\right)^2\,\right\}^{1/2}\\
\no &\leq& \left(\,\sum_k |c_k|^2\,\zeta_k\,\right)^{1/2}\,\left\{\,\sum_k \frac{1}{\zeta_k}
\left(\,\sum_j\,|c_j|^2\,\zeta_j\,\right)\,\left(\,\sum_j\,\frac{|B_{j,k}|^2}{
\zeta_j}\,\right)\,\right\}^{1/2}\\
\no &\leq& \Emax\,\left\{\,\sum_{j,k} \frac{|B_{j,k}|^2}{\zeta_j\,\zeta_k}\,\right\}^{1/2}.\\
&&
\ear

For the proof of the Corollary, note that:

\bar
\no |<\frac{\partial \psi_j}{\partial X}|\psi_k>| &\leq& <|\frac{\partial \psi_j}{
\partial X}|^2>^{1/2}\,<|\psi_k|^2>^{1/2}\\
\no &\=&  <|\frac{\partial \psi_j}{
\partial X}|^2>^{1/2}\\
\no &\=& <\psi_j| - \frac{\partial^2 \psi_j}{\partial X^2}>.\\
&&
\ear

\ni We know from the energy bound

\be 
\frac{\hbar^2}{2M}\,<\psi| - \frac{\partial^2 \psi}{\partial X^2}> \ph \leq \ph \Emax;
\ee

\ni therefore

\be 
|<\frac{\partial \psi_j}{\partial X}|\psi_k>| \ph \leq \ph \frac{\sqrt{2M\Emax}}{\hbar},
\label{Ebound}
\ee

\ni from which we derive a uniform upper bound on the $B_{j,k}$; the Corollary follows.

The reason to doubt that the hypothesis of the Corollary holds in the 1DTM is that
the upper and lower bounds produced in section \ref{1dtm} suggest that the energies have the form
of sums of squares of integers. But reflecting on the bound:

\be
\int_1^{\infty}\,dx_1 ...\int_1^{\infty}\,dx_N\,
\left\{\,\frac{1}{x_1^2+x_2^2 + ... +x_N^2}
\,\right\} \ph \leq \ph \sum_{k_1 = 1,...,k_N=1}^{\infty}\,\left\{\,\frac{1}{k_1^2 + ... + k_N^2}
\,\right\},
\ee

\ni and making a substitution in the integral using generalized polar coordinates
(e.g., for $N=2$, $x_1 = r\cos(\theta) + 1$ and $x_2 = r\sin(\theta) + 1$)
one sees that the multiple integral will contain an integration over $r$ of form:

\be
\int^{\infty}\,dr\,r^{N-1}\,\frac{1}{r^2} \= \infty,
\ee

\ni whenever $N > 1$. Hence

\be
\sum_{k_1 = 1,...,k_N=1}^{\infty}\,\left\{\,\frac{1}{k_1^2 + ... + k_N^2} 
\,\right\} \= \infty.
\ee

\end{document}